# Evaluation of concept drift adaptation for acoustic scene classifier based on Kernel Density Drift Detection and Combine Merge Gaussian Mixture Model

Ibnu Daqiqil Id, Masanobu Abe, Sunao Hara (Okayama University)

## 1 Introduction

Usually, in Machine learning and predictive analysis, the training and testing data is a random variable generated independently from an underlying stationary distribution[1]. However, in many applications, especially in acoustic scene classification, the system being examined with different data distribution due to several reasons like the change of user's behavior or environment, non-stationary noise, diverse sound events, overlapping in time or frequency audio events, and echo or reverberant operating conditions. The situation where the data distribution change is known as concept drift[2]. As a result, this situation causes the models to undergo performance degradation[3].

In [4], we proposed a Combine Merge Gaussian mixture model (CMGMM) and Kernel Density Estimation (KD3) as a method to detect and adapt to the concept drift problem in the acoustic scene classification problem. CMGMM is an algorithm based on Gaussian Mixture Model that can adapt concept drift by add or modify components to accommodate the emerging concept drift.

KD3 is a concept drift detection algorithm based on the kernel density estimation (KDE) method to estimate the probability density function with one random variable. By comparing two density functions, it is possible to establish the degree of variation in values between the corresponding variables—the greater the variation, the more evidence for the concept drifts. In addition to detecting concept drifts, KD3 also acts as a data collector for the adaptation process by marking a warning zone where data begin to show concept drift symptoms.

In [4], we showed that the KD3 plays an essential role in the model adaptation. This algorithm required three hyperparameters, namely α, β, and $\hbar$. Hyperparameter α is the margin of windows variation to detect concept drift, β is the margin to accumulate the windows variation over time, and $\hbar$ is the window length. In [4], a parameter grid search was performed to find the optimal hyperparameter. The window length $\hbar$ was set to 45, the drift margin α was set to 0.1, and the warning margin β was set to 0.0001. Those hyperparameters are obtained from datasets containing random concepts drift in three scenarios.

Based on the experiment result, hyperparameters α can significantly affect the performance of the model. Reducing α means making KD3 more sensitive to changes. A sensitive KD3 could lead to an overfitting problem when the number of components grows significantly, then redundant mixture components located close together are introduced. Overfitting can have an adverse impact on predictions, then degrade model performance.

This paper aims to improve and evaluate the CMGMM and KD3 accuracy by tuning the hyperparameter in several concept drift types and scenarios.

## 2 The component pruning algorithm

CMGMM tends to increase the number of components because it works by merging two mixture models (current and adapted models). This mechanism led to an overfitting problem as the frequency of adaptation increases due to the sensitive KD3 hyperparameter.

As a solution to this problem, we propose a component reduction mechanism based on weight ($w$) and covariance (Σ) in CMGMM. These reduced components were identified by the ratio of $w^2$ and $\Sigma^2$ being close to zero. In practice, components with $w$ very close to zero are ignored by the model, and components with large covariance tend to overlap with other components.

## 3 Experiment

### 3.1 Experiment Setup

This experiment will evaluate CMGMM and KD3 accuracy systematically using a combination of hyperparameters α from 0.1 to 0.001, β from 0.0001 to 0.000001, and $\hbar$ from 45 to 300. Furthermore, we will also evaluate the component pruning mechanism's performance using prequential or interleaved-test-then-train with the maximum number of instances for the test or train is 100.

### 3.2 Dataset

The dataset for training and evaluating the KD3 and CMGMM was generated by selecting 15 types of scenes from the TUT Acoustic Scenes 2017 dataset [5] and TAU Urban Acoustic Scenes 2019 dataset [6]. Then we generate four types of concepts drifted datasets in three event audio placement scenarios, namely T1, T2, and T3. In T1, several types of unique event sounds related to each scene were overlaid several times with random timing and gains. In scenario T2, several sounds were randomly selected from a particular group of event sounds, making the concept drift of this scenario to be more complicated than that of scenario T1. Then scenario T3 was quite similar to scenario T2. The difference was that a group of event sounds could exist in several scenes. The overall dataset consisted of 12,000 audio segments of ten seconds each, equally distributed between 15 different scenes and annotated with their ground-truth labels.

As shown in Fig. 1, the types of concept drift used to generate the dataset are:

- Simple Concept Drift (A) is characterized by a new concept that appears at a particular time point, and then at a particular time in the future, the concept is disappear or replaced by another novel concept.
- Gradual concept drift (B) is characterized by a new concept that appears at a particular time point, and then at a particular time in the future, another novel concept adds the concept in the audio scene.
- Recurring concept drift type 1 (C1) is characterized by cyclical emerging concept change, or a previous concept reappeared after some time.
- Recurring concept drift type 2 (C2) is a combination of Recurring concept drift type 1 and Gradual concept drift. The characteristic of this type of drift is cyclical emerging concept change and the stack of a new emerging concept.

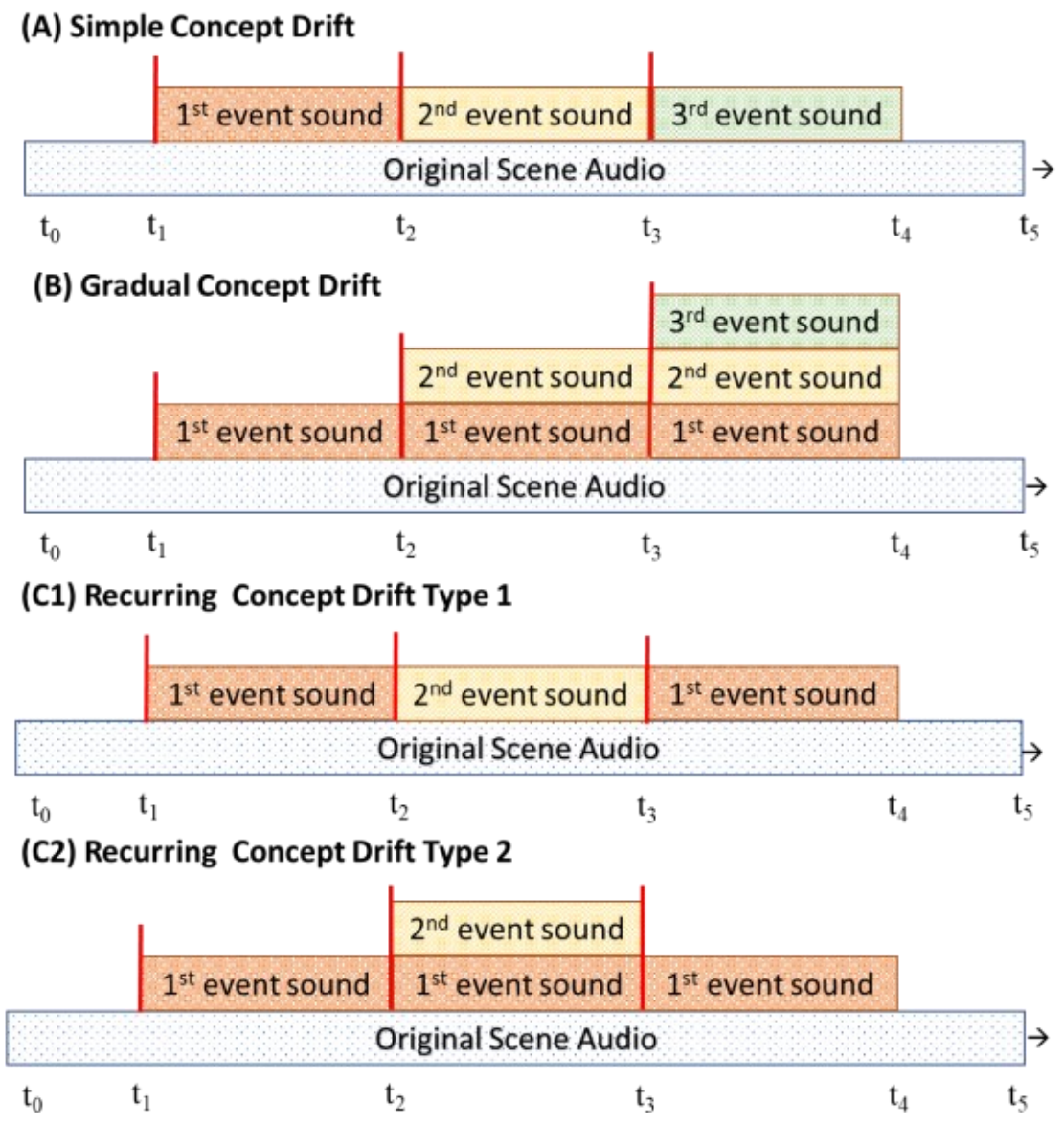

Figure 1. Type of Concept Drift

## 3.3 Experiment Result

### 3.3.1 Effects of hyperparameters β and ɦ

Figure 2 shows the average accuracy in all concept drift types, according to hyperparameters β and ɦ. Judging from the results, we found that hyperparameters β and ɦ have no significant effect on accuracy. However, hyperparameter α significantly affects performance. Therefore, in the rest of the paper, we will discuss the experiment results for hyperparameter α. In those experiments, β and ɦ are set at 0.001 and 45, respectively.

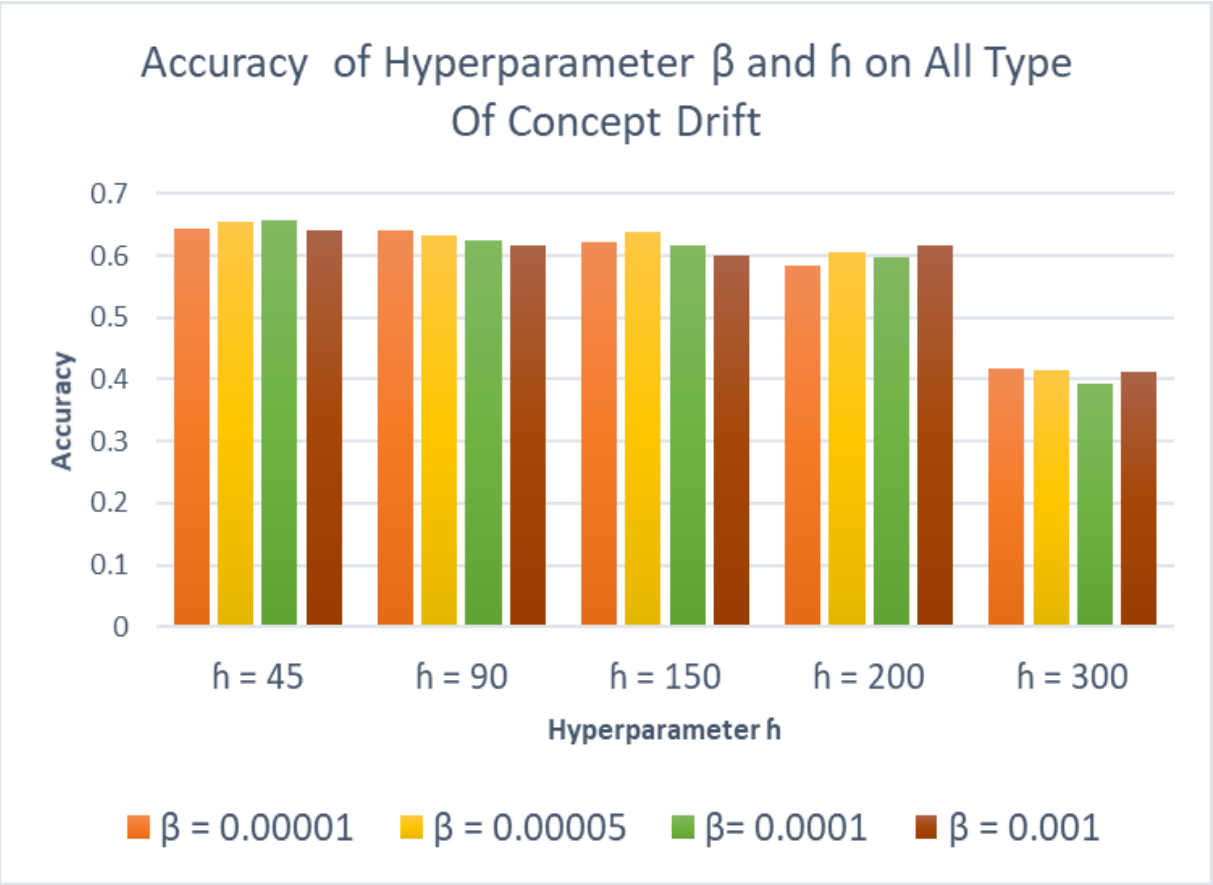

Figure 2. Result of hyperparameter β and ɦ in four types of concept drift

### 3.3.2 Effects of hyperparameters α

Table 1 lists the experimental results for four different types of concept drift and event audio placement scenario. In general, each concept drift type has its respective hyperparameter according to concept drift characteristics. For example, simple concept drift and Gradual concept drift have similar characteristics where there are no repeating concepts, so that the hyperparameter patterns they have are also similar. Furthermore, how the audio event is placed has no significant effect. For example, the simple concept drift hyperparameter on T1 is the same as the hyperparameter for T2 and T3.

In simple concept drift, a model with hyperparameter α = 0.05 shows the best results. The value of α = 0.05 is a quite sensitive hyperparameter value because the number of concept drifts detected increases significantly. If the hyperparameter α is increased, the frequency of model adaptation decrease Therefore, by accelerating the update frequency, the loss received by the model is reduced. This pattern also occurs in the T1, T2, and T3 scenarios. Furthermore, in cases where new concepts are replaced while old concepts do not reappear, the model must increase the model adaptation frequency.

Gradual concept drift requires a more sensitive hyperparameter than a simple concept drift hyperparameter because the emerging concept is gradually increasing. Therefore, reducing α to increase the adaptation frequency would increase model performance. The optimal hyperparameters are α = 0.01. Although T1 accuracy is not the best performance in this type of concept drift, the average accuracy is higher, and the number of adaptations is better than α = 0.005.

Table 1. Accuracy and number of model adaptation result for four types of concept drift and scenario using β=0.001 and ɦ=45

| α | Accuracy | | | Num. of Model Adaptation | | |
|---|---|---|---|---|---|---|
| | T1 | T2 | T3 | T1 | T2 | T3 |
| **(A) Simple Concept Drift Dataset** | | | | | | |
| α=0.1 | 0.6568 | 0.6333 | 0.6910 | 28 | 32 | 37 |
| **α=0.05** | **0.7413** | **0.7064** | **0.6978** | 55 | 54 | 55 |
| α=0.01 | 0.7069 | 0.6973 | 0.6767 | 135 | 133 | 141 |
| α=0.005 | 0.7137 | 0.7007 | 0.6842 | 157 | 175 | 152 |
| α=0.001 | 0.7066 | 0.677 | 0.6524 | 196 | 186 | 201 |
| **(B) Gradual Concept Drift Dataset** | | | | | | |
| α=0.1 | 0.6007 | 0.5620 | 0.5831 | 35 | 37 | 35 |
| α=0.05 | 0.6173 | 0.5863 | 0.5818 | 51 | 60 | 52 |
| **α=0.01** | 0.6922 | **0.6187** | **0.6238** | 124 | 128 | 130 |
| α=0.005 | **0.7050** | 0.5448 | 0.5837 | 157 | 123 | 152 |
| α=0.001 | 0.7002 | 0.5996 | 0.5996 | 168 | 165 | 169 |
| **(C1) Recurring Concept Drift Dataset type 1** | | | | | | |
| **α=0.1** | **0.8792** | **0.8836** | **0.8705** | 35 | 34 | 39 |
| α=0.05 | 0.8432 | 0.8511 | 0.8373 | 67 | 68 | 67 |
| α=0.01 | 0.7458 | 0.7661 | 0.7723 | 181 | 166 | 168 |
| α=0.005 | 0.7254 | 0.7373 | 0.7483 | 182 | 185 | 180 |
| α=0.001 | 0.7127 | 0.724 | 0.7148 | 206 | 202 | 207 |
| **(C2) Recurring Concept Drift Dataset type 2** | | | | | | |
| **α=0.1** | **0.8668** | **0.866** | **0.87** | 37 | 38 | 37 |
| α=0.05 | 0.8433 | 0.826 | 0.8288 | 65 | 73 | 74 |
| α=0.01 | 0.7628 | 0.765 | 0.744 | 174 | 164 | 170 |
| α=0.005 | 0.7490 | 0.721 | 0.7249 | 198 | 179 | 194 |
| α=0.001 | 0.7323 | 0.716 | 0.7213 | 207 | 203 | 202 |

Recurring concept drift type 1 is different from simple and gradual concept drift patterns. In this type of concept drift, small α that is sensitive to change tends to have poor performance. The experimental results also show the optimal hyperparameter values that used in [4] is similar (α = 0.1, β = 0.001 and ɦ = 45). One of the advantages of this hyperparameter is the large difference between α and β could make a larger warning zone, so the model has enough data for adaptation. The results obtained in this experiment are also in line with the results obtained in previous studies where the dataset used contains a random pattern. In random pattern concept drift, there is a possibility that new concepts that emerge will likely repeat themselves in the future.

The Recurring concept drift type 2 is a combination of recurring and gradual concept drift, which still has the same pattern as recurring concept drift type 1. In this type, the best results are also shown on α = 0.1, then the performance decrement also happens if we decrease the hyperparameter α.

In all type of concept drift, if the α value is reduced less than 0.01, the number of components increase significantly. Table 2 shows on in the initial training ($E_1$), the average of component number for simple concept drift (A) is 5.67 per scene. However, at the end of the experiment($E_n$), the average component number had increased four times to 21.8 per scene. This significant growth in the number of components affects the model's performance because too many components of the GMM are the cause of overfitting problems.

Table 2. Number of component at first training($E_1$) and last adaptation ($E_n$) in four type of concept drift using α = 0.001 in T1 scenario

| Scene Label | A | | B | | C1 | | C2 | |
|---|---|---|---|---|---|---|---|---|
| | $E_1$ | $E_n$ | $E_1$ | $E_n$ | $E_1$ | $E_n$ | $E_1$ | $E_n$ |
| Airport | 4 | 20 | 4 | 21 | 4 | 18 | 4 | 18 |
| Beach | 9 | 34 | 8 | 31 | 10 | 35 | 11 | 31 |
| Bus | 4 | 23 | 4 | 23 | 4 | 15 | 4 | 17 |
| Cafe/restaurant | 14 | 34 | 12 | 36 | 8 | 18 | 8 | 17 |
| City center | 4 | 18 | 4 | 20 | 4 | 15 | 4 | 16 |
| Grocery store | 4 | 17 | 7 | 21 | 4 | 16 | 4 | 15 |
| Home | 10 | 35 | 10 | 34 | 12 | 34 | 12 | 28 |
| Metro | 4 | 18 | 4 | 19 | 4 | 12 | 4 | 12 |
| Office | 8 | 29 | 8 | 39 | 6 | 15 | 6 | 15 |
| Public Square | 4 | 17 | 4 | 25 | 4 | 12 | 4 | 11 |
| Residential Area | 4 | 15 | 4 | 22 | 4 | 13 | 4 | 10 |
| Shopping Mall | 4 | 14 | 4 | 24 | 4 | 16 | 4 | 16 |
| Street Pedestrian | 4 | 18 | 4 | 22 | 4 | 14 | 4 | 15 |
| Street Traffic | 4 | 17 | 4 | 23 | 4 | 13 | 4 | 14 |
| Tram | 4 | 18 | 4 | 23 | 4 | 11 | 4 | 12 |

### 3.3.3 Effects of the component pruning algorithm

It can be noticed in figure 3, implementation of the components pruning to overcome the overfitting problem shows better results in almost all types of drift concepts and scenarios. In simple concept drifts show an increase in T1 of 0.99%, T2 of 0.7 %, and T3 of 0.4%. Gradual concept drift show an increase in T1 of 5.55%, T2 of 1.29%, and T3 of 0.4%. Furthermore, in the model that requires a high-frequency model adaptation like simple concept drift and gradual concept drift, component pruning could control the active component numbers and improve the accuracy. However, there was a decrease in T2 of recurring concept drift type 1 by 1.27% and T1 of recurring concept drift type 2 by 0.18%.

Figure 4 shows detailed results of simple concept drift in the T1 scenario in mean and window accuracy. Window accuracy is the accuracy measured in each prequential batch of 100 data. In this figure, it is clear that there is an increase in performance in certain ranges when using pruning components. For example, from 1700 to 2900, the window accuracy of CMGMM with component pruning shows better performance than CMGMM.

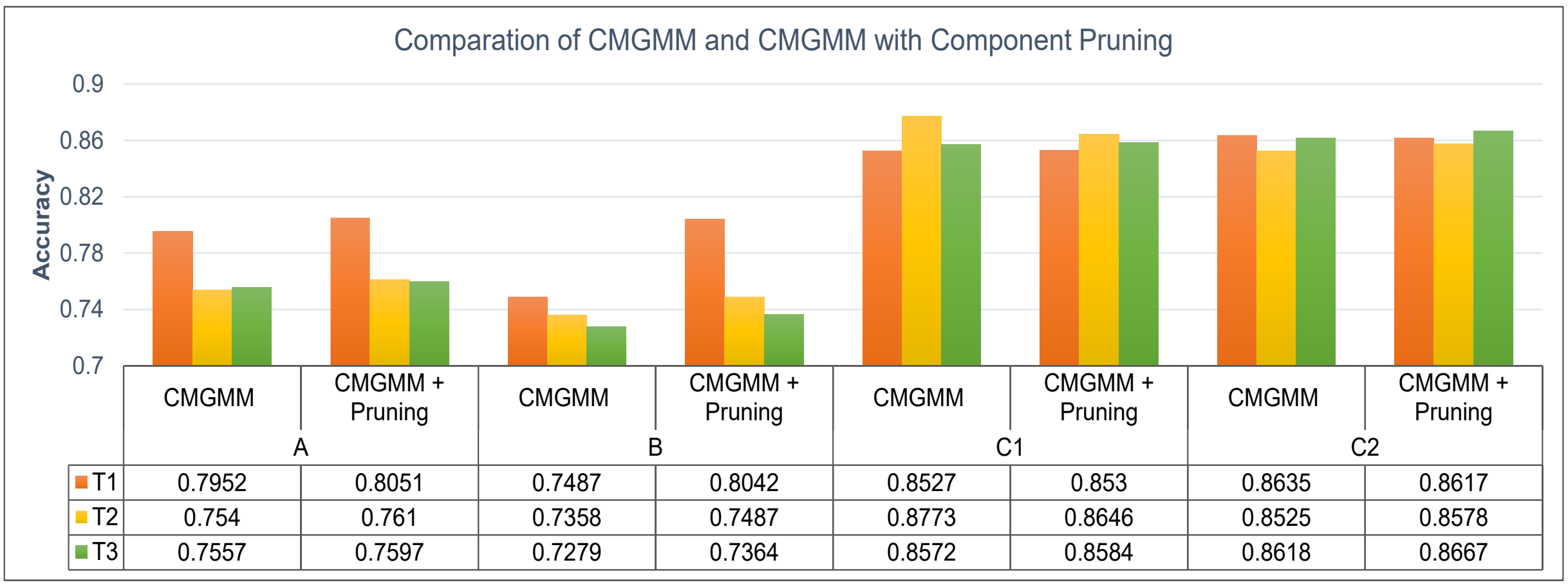

| | A: CMGMM | A: CMGMM + Pruning | B: CMGMM | B: CMGMM + Pruning | C1: CMGMM | C1: CMGMM + Pruning | C2: CMGMM | C2: CMGMM + Pruning |
|---|---|---|---|---|---|---|---|---|
| T1 | 0.7952 | 0.8051 | 0.7487 | 0.8042 | 0.8527 | 0.853 | 0.8635 | 0.8617 |
| T2 | 0.754 | 0.761 | 0.7358 | 0.7487 | 0.8773 | 0.8646 | 0.8525 | 0.8578 |
| T3 | 0.7557 | 0.7597 | 0.7279 | 0.7364 | 0.8572 | 0.8584 | 0.8618 | 0.8667 |

Figure 3. Performance comparison of CMGMM with component pruning and without component pruning in four type of concept drift

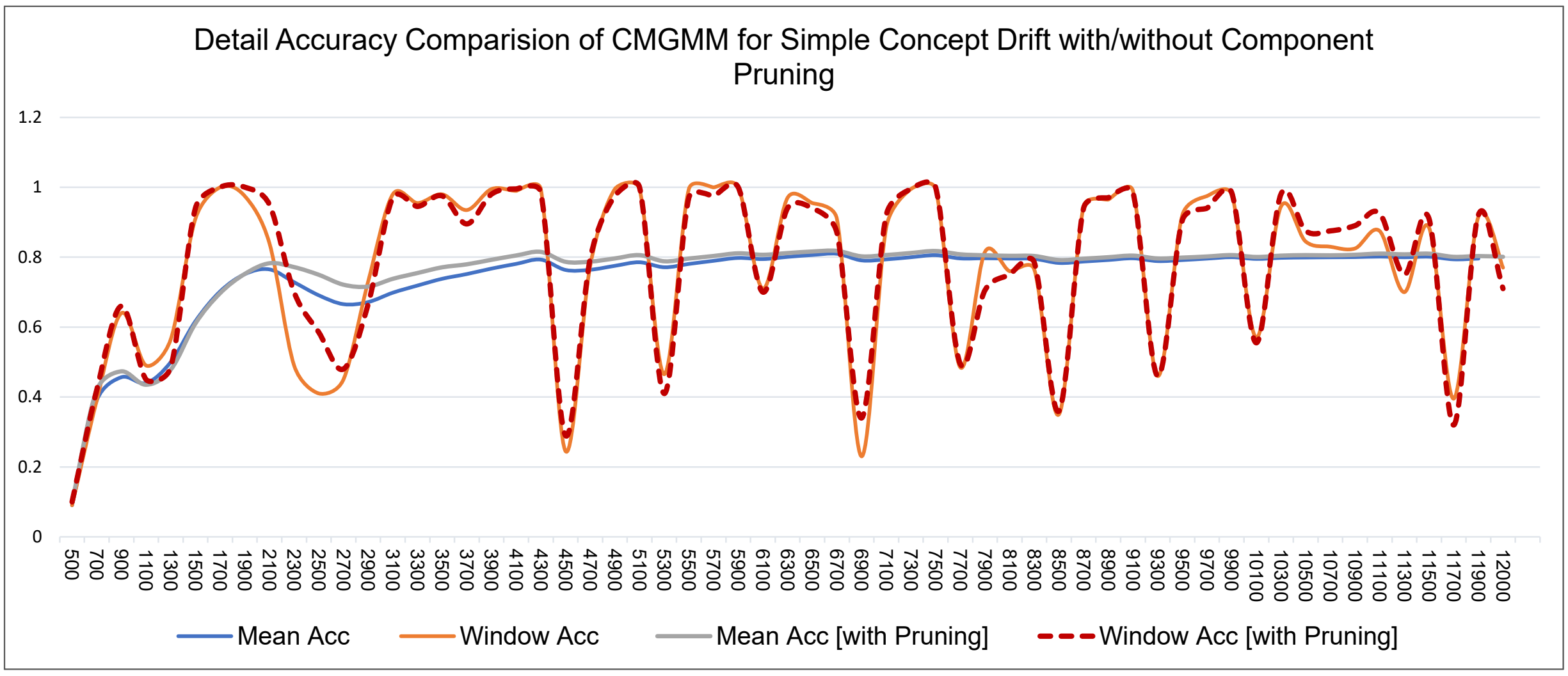

Figure 4. Detailed result of the component pruning mechanism on Simple Concept Drift in T1 Scenario

## 4 Conclusion

Based on the experimental results, all concepts drift types have their respective hyperparameter configurations. Simple and gradual concept drift have similar pattern which requires a smaller α value than recurring concept drift because, in this type of drift, a new concept appear continuously, so it needs a high-frequency model adaptation. However, in recurring concepts, the new concept may repeat in the future, so the lower frequency adaptation is better. Furthermore, high-frequency model adaptation could lead to an overfitting problem. Implementing CMGMM component pruning mechanism help to control the number of the active component and improve model performance.